\documentclass{article}
\usepackage{amsmath}
\usepackage[doublespacing]{setspace}
\usepackage[a4paper,includemp,reversemp,pdftex,twocolumn=false,nohead]{geometry}

\input{tcilatex}
\begin{document}

\title{On stability in dynamical Prisoner's dilemma game with non-uniform
interaction rates}
\date{}
\author{M.I.Shehata \\
%EndAName
Mathematics Department, Faculty of Science, Mansoura 35516, EGYPT\\
E-mail: mishehata@yahoo.com}
\maketitle

\begin{abstract}
Stability of evolutionary dynamics of non-repeated Prisoner's Dilemma game
with non-uniform interaction rates [1], via benefit and cost dilemma is
studied . Moreover , the stability condition $(\frac{b+c}{b-c})^{2}<$ $%
r_{1}r_{3}$ \ is derived in case of coexistence between cooperators and
defectors .If $r_{1},r_{3}\longrightarrow \infty $ cooperation is the
dominant strategy and defectors can no longer exploit cooperators.

\end{abstract}

\begin{description}
\item[{\protect\large Introduction}] 
\end{description}

Evolutionary game theory[3] was first introduced by John Maynard Smith in
1973[2]. He invented the important concept of an evolutionarily stable
strategy (ESS) that resist invasion of other strategies in infinitely large
populations. Prisoner's dilemma (PD) game[4] studies cooperation between
unrelated individuals. In non-repeated PD,individuals are either cooperators
or defectors, acting accordingly whenever two of them interact. They both
receive R upon mutual cooperation and P upon mutual defection. A defector
exploiting a cooperator gets an amount T and the exploited cooperator
receives S, such that T\TEXTsymbol{>}R\TEXTsymbol{>}P\TEXTsymbol{>}S .So, it
is best to defect, regardless of the opponent's decision, which in turns
makes cooperators unable to resist invasion by defectors.

In this paper, we study the effect of non-uniform interaction rates on
evolutionary dynamics of non-repeated PD game [1], via benefit and cost
dilemma . In section 1, we present the classical approach of the replicator
equation of PD with uniform interaction rate between any two individuals.
This rate does not depend on the strategies of these individuals. In section
2,we assume that the interaction rates are not uniform. For example, players
who use the same strategy might interact more frequently than players who
use different strategies. Non-uniform interaction rates lead to non-linear
fitness functions and therefore allow richer dynamics than the classical
replicator equation, which is based on linear fitness functions. If strategy
D is a strict Nash equilibrium[5], then it remains uninvadable for positive
non-uniform interaction rates. If D dominates C then non-uniform interaction
rates can introduce a pair of interior equilibria; one of them is stable the
other one is unstable. If C and D coexist, then non-uniform \ interaction
rates cannot change the qualitative dynamics, but alter the location of the
stable equilibrium.

\section{Dynamical PD game with uniform interaction rate}

Consider PD game between two players each having two strategies
C(cooperation) and D (defection) and Cooperator pay a benefit $b$ at a cost $%
c$ to his Cooperator opponent, whereas Defector opponent only receive a
benefit $b$ [6]. Then we have payoff matrix

\begin{equation}
\left[ 
\begin{array}{cc}
b-c & -c \\ 
b & 0%
\end{array}%
\right]
\end{equation}%
where, $b>c>0.$We denote by $x$ and $y$ the frequency of individuals
adopting strategy C and D, respectively. We have $x+y=1.$

In case of uniform interaction rate between any two individuals. We assume
that interaction rate is independent\ of their strategies. The selection
dynamics can be described by replicator equation [7]

\begin{equation}
\dot{x}=x(1-x)(f_{C}-f_{D})
\end{equation}%
The fitness of $C$ and $D$ players are linear functions of $x$ and $y$,
given by

\begin{equation}
f_{C}=(b-c)x-cy\text{ \ },\text{\ }f_{D}=bx.
\end{equation}

The entire population will eventually consist of Defectors ,since $b>b-c$
and $0>-c$ .The only stable equilibrium is $\ x=0.$ D is a strict Nash
equilibrium,and therefore an evolutionarily stable strategy (ESS).

\section{Stability in dynamical PD game with non-uniform interaction rates}

Suppose that \ an interaction between two players depends on their
strategies. Let $r_{1},r_{2},r_{3}$ \ be reaction rates between each two
players and $r_{1},r_{2},r_{3}>0$ . where,$r_{1}$ is\ an interaction rate of
Cooperator with another Cooperator, $r_{2}$ is\ an interaction rate of
Cooperator with Defector, and $r_{3}$ is\ an interaction rate of Defector
with another Defector.

For uniform interaction rates $r_{1}=r_{2}=r_{3}$ \ the strategy D dominates
the strategy C. Hence, eventually the entire population will consist of
defectors.

However, if players only interact with opponents of the same strategy, then
cooperators cannot be exploited by defectors. In this case, where $r_{2}=0$
and $\ r_{1},r_{3}>0$ ,cooperation is the dominant strategy, because $b-c>0.$%
Assume $r_{2}\neq 0$ which means that cooperators and defectors do interact.
Without loss of generality, assume that $r_{2}=1$.

The fitness of individuals is determined by the average payoff over a large
number of interactions. Therefore, the fitness of C and D players are
non-linear functions of $x$ and $y$, given by

\begin{equation}
f_{C}=\frac{(b-c)r_{1}x-cy}{r_{1}x+y}\text{ \ \ },\text{\ \ \ \ }f_{D}=\frac{%
bx}{x+r_{3}y}
\end{equation}%
The replicator equations can be reduced to

\begin{equation}
\dot{x}=x(1-x)(f_{C}-f_{D})
\end{equation}%
where,

\begin{equation*}
f_{C}-f_{D}=\frac{((b-c)r_{1}x-cy)(x+r_{3}y)-(bx)(r_{1}x+y)}{%
(r_{1}x+y)(x+r_{3}y)}
\end{equation*}%
The equilibrium points are either on the boundary or in the interior. The
boundary points $x=0$ is stable equilibrium while $x=1$ is unstable
equilibrium. If $f_{C}-f_{D}=0$ and $x+y=1$then we have two interior
equilibrium are given by

\begin{equation}
x^{\ast }=\frac{-(\alpha -2\gamma )\pm \sqrt{\alpha ^{2}-4\beta \gamma }}{%
2(\beta +\gamma -\alpha )}
\end{equation}%
where,

\begin{equation*}
\alpha =r_{1}r_{3}(b-c)-(b+c)\text{ \ \ , \ }\beta =-r_{1}c\text{ \ \ \ \ ,
\ \ }\gamma =-r_{3}c\text{\ }
\end{equation*}

\begin{equation*}
\alpha ^{2}>4\beta \gamma \text{ \ \ , \ \ }\beta +\gamma <\alpha \text{ \ \
\ , \ \ }\alpha >2\gamma \text{ \ \ , \ \ }\alpha >2\beta
\end{equation*}

In case of non- uniform interaction rate ,the selection dynamics depend on
the size of $r_{1}r_{3}$ relative to $\rho ^{2}$.[1] where,

\begin{equation}
\rho =\frac{b+c}{b-c}\text{ \ and \ }\rho >1\text{ .}
\end{equation}%
If $r_{1}r_{3}$ $<(\frac{b+c}{b-c})^{2}$ then defection is the dominant
strategy. When $r_{1}r_{3}$ $=(\frac{b+c}{b-c})^{2}.$we have the bifurcation
[8] point at

\begin{equation}
x^{\ast }=\frac{\sqrt{r_{3}}}{\sqrt{r_{1}}+\sqrt{r_{3}}}
\end{equation}%
\ If $r_{1}r_{3}$ $>(\frac{b+c}{b-c})^{2}$ the two interior equilibria (6)
move symmetrically away from the bifurcation point (8) toward the end points.

As $r_{1}$ and $r_{3}$ \ increase the\ stable interior equilibrium point
moves from the bifurcation point (8) closer toward 1.Whereas, \ unstable
interior equilibrium point moves from the bifurcation point (8) closer
toward 0.As $r_{1},r_{3}\longrightarrow \infty $ we recover the case where $%
r_{2}=0$ and cooperation is the dominant strategy as defectors can no longer
exploit cooperators.

\section{Conclusion}

For the non-repeated PD game, coexistence between cooperators and defectors
is possible if the ratio of homogeneous $r_{1},r_{3}$over heterogeneous $%
r_{2}$interaction rates exceeds a critical value.\ If $r_{1}=r_{3}=\rho $,
then the pair of equilibria approaches cooperator frequency of $x=\frac{1}{2}
$ which is entirely independent of the payoff matrix, as long as $b>b-c>0>-c$%
\textit{\ .} Both equilibria are stable, one consists of defectors alone,
and the other consists of a mixture of defectors and cooperators.

\end{document}